\documentclass[prb,aps,twocolumn,amsfonts,amsmath,amssymb,showkeys,showpacs]{revtex4}

\usepackage{amsmath}          
\usepackage[final]{graphicx}   
\usepackage{verbatim}           
\usepackage{color}                 
\usepackage{subfigure}          
\usepackage{hyperref}           

\let\omp\marginpar\relax
\def\marginpar#1{\omp{\color{red}#1}}
\newlength\figurewidth
\AtBeginDocument{\setlength\figurewidth{0.98\linewidth}}

\def\Ang{\AA}
\def\Vshear{\AA/ps}
\def\aSi{{\it a}-Si}
\def\cSi{{\it c}-Si}
\def\eal{\textit {et al.}}

\begin{document}

\title{Crystallization of amorphous silicon induced by mechanical shear deformations}

\author{Ali Kerrache}                           %
\email{ali.kerrache@umontreal.ca}     %
\affiliation{D\'epartement de Physique, Regroupement Qu\'eb\'ecois sur les Mat\'eriaux de Pointe, Universit\'e de Montr\'eal, C.P. 6128, succ. Centre-ville, Montr\'eal (Qu\'ebec) H3C 3J7, Canada.}

\author{Normand Mousseau}                       %
\email{normand.mousseau@umontreal.ca} %
\affiliation{D\'epartement de Physique, Regroupement Qu\'eb\'ecois sur les Mat\'eriaux de Pointe, Universit\'e de Montr\'eal, C.P. 6128, succ. Centre-ville, Montr\'eal (Qu\'ebec) H3C 3J7, Canada.}

\author{Laurent J. Lewis}                   %
\email{laurent.lewis@umontreal.ca}    %
\affiliation{D\'epartement de Physique, Regroupement Qu\'eb\'ecois sur les Mat\'eriaux de Pointe, Universit\'e de Montr\'eal, C.P. 6128, succ. Centre-ville, Montr\'eal (Qu\'ebec) H3C 3J7, Canada.}

\date{\today}

\begin{abstract}

We have investigated the response of amorphous silicon ({\it a}-Si), in particular crystallization, to external mechanical shear deformations using classical molecular dynamics (MD) simulations and the empirical Environment Dependent Inter-atomic Potential (EDIP) [Phys. Rev. B {\bf 56}, 8542 (1997)]. In agreement with previous results we find that, at low shear velocity and low temperature, shear deformations increase disorder and defect density. At high temperatures, however, the deformations are found to induce crystallization, demonstrating a dynamical transition associated with both shear rate and temperature. The properties of {\it a}-Si under shear deformations and the extent at which the system crystallizes are analyzed in terms of the potential energy difference (PED) between the sheared and non-sheared material, as well as the fraction of defects and the number of particles that possess a crystalline environment.

\end{abstract}


\maketitle

\section{\label{intro}Introduction}

Amorphous silicon (\aSi) is widely used in the electronic and semiconductor industries. From a theoretical point of view, because of its simplicity, it has become the archetype of covalent amorphous systems that include silica and chalcogenide glasses. While these materials have been extensively studied for almost four decades, their behavior under external forces has received significant attention only recently; \cite{Falk, Maloney, Ivashchenko, Rottler, Argon, Demkowicz, Talati, Stich, Car} yet, many questions remain unanswered. For example, while the plasticity of \aSi~is attributed to the presence of liquid-like particles associated with 5-fold coordinated atoms, \cite{Argon, Demkowicz} a complete picture of the deformation effects on disordered and amorphous materials is far from being complete. It has been shown that, under external forces, amorphous materials may crystallize,\cite{Chen, Vleeshouwers, Tarumi, Pelaz} providing insight into the nucleation process for these covalent materials which is only partially understood. \cite{Robertson, Nakhmanson} Indeed, crystallization is often studied using amorphous-crystal or liquid-crystal interfaces\cite{Krzeminski, Buta, Brambilla, Kerrache08} and this is in particular the case for Si.\cite {Krzeminski, Buta, Bernstein98, Mattoni, Lewis96} Crystallization may also result from the application of external forces such as mechanical shear deformations\cite{Duff, Mokshin-08, Mokshin-09} or magnetic fields.\cite{Park}

The properties of bulk crystalline silicon (\cSi) have been investigated by means of molecular dynamics (MD) simulations using both classical potentials\cite{Argon, Demkowicz} and \emph{ab-initio} methods.\cite{Godet} It has been demonstrated that, under the effect of shear, the interatomic bonds lose their covalent character until a metallic state is established in the shear direction;\cite{Godet} upon inverting the deformation, however, the perfect diamond structure is recovered. In the case of disordered materials, it has been shown that small strains bring these systems to deeper energy minima in the glassy state.\cite{Lacks} Under deformations, these materials might escape the minima through the high energy barrier and visit other minima of the energy landscape giving rise to new local rearrangements of the amorphous structure.\cite{McKenna} These features depend on strain, shear rate and temperature at which the deformations are applied.\cite{Ivashchenko, Talati, Kerrache}

Strain can also play a role in re-ordering the network. Lee~\eal,\cite{Lee} for example, have investigated deformation and grain growth in partially crystallized nickel by means of MD simulations and showed that shear deformations can enhance crystallization in amorphous materials. Recently, Mokshin and Barrat\cite{Mokshin-08, Mokshin-09} have demonstrated that shearing an initially amorphous system leads to an increase of the nano-crystalline ordering. They examined both a one-component Lennard-Jones system ~\cite{Mokshin-08} and the one-component short-range Dzugutov potential.\cite{Dzugutov} As the disordered phase of these single-component close-packed systems is known to be very unstable (even though the Dzugutov model does better in this respect~\cite{Mokshin-09}), however, it is not clear whether this shear-induced crystallization is generic. Talati~\eal,\cite{Talati} using both the Tersoff\cite{Tersoff} potential and various versions of the Stillinger-Weber\cite{SW} potential for silicon, found no evidence of ordering from the amorphous state in the range of shear velocities and temperatures they considered. The crystallization of computer-simulated silicon is, however, known to be potential dependent,\cite{Krzeminski} and both standard Tersoff and the Stillinger-Weber potentials tend to over-stabilize the liquid-like environments in the amorphous phase. Given the importance of shear-induced crystallization, further study of this covalent system is clearly warranted.

In view of this situation, we have carried out a detailed MD study of the properties of \aSi~under low-velocity shear deformations, describing the energetics of the silicon atoms in terms of the Environment Dependent Inter-atomic Potential (EDIP).\cite{Bazant, Justo} The present study follows up on a previous investigation of the effects of shear velocity and temperature on amorphous silicon.\cite{Kerrache} In particular, it deals with the extent of the distribution of defects when a low shear velocity is imposed on the system or when the deformations are applied at high temperature. Our previous study was concerned with the effects of shear velocity and temperature on amorphous silicon; it was found that the impact of an externally applied strain can be almost fully compensated by increasing the temperature, allowing the system to respond more rapidly to the deformation. In this article we show that that, depending on shear velocity and temperature, \aSi~can either crystallize or remain in its glassy state, following a more complex kinetic path than the Lennard-Jones system but consistent with the ''universality'' proposed by Mokshin and Barrat.\cite{Mokshin-08, Mokshin-09}


\section{\label{model}Computer Model}

As mentioned above, our calculations were performed within the framework of MD and the EDIP classical potential.\cite{Bazant, Justo} This potential, which was fitted to various zero-temperature bulk phases and defect structures, reproduces accurately the structure and the dynamics of \aSi~as well as different crystalline structures. It possesses remarkable transferability for zero-temperature properties, including elastic constants, bulk crystal structures and point defects.\cite{Justo, Allred} This potential has been used with success to study ion-beam induced amorphization of \cSi~\cite{Caturla, Nord} as well as crystallization of \aSi~at the amorphous-crystalline silicon interface.\cite{Buta, Bernstein98} All simulations were performed using the massively parallel MD package LAMMPS developed by \emph{Sandia National Laboratories}.\cite{LAMMPS}

\begin{figure}[ht!]
\includegraphics[width=0.9\figurewidth]{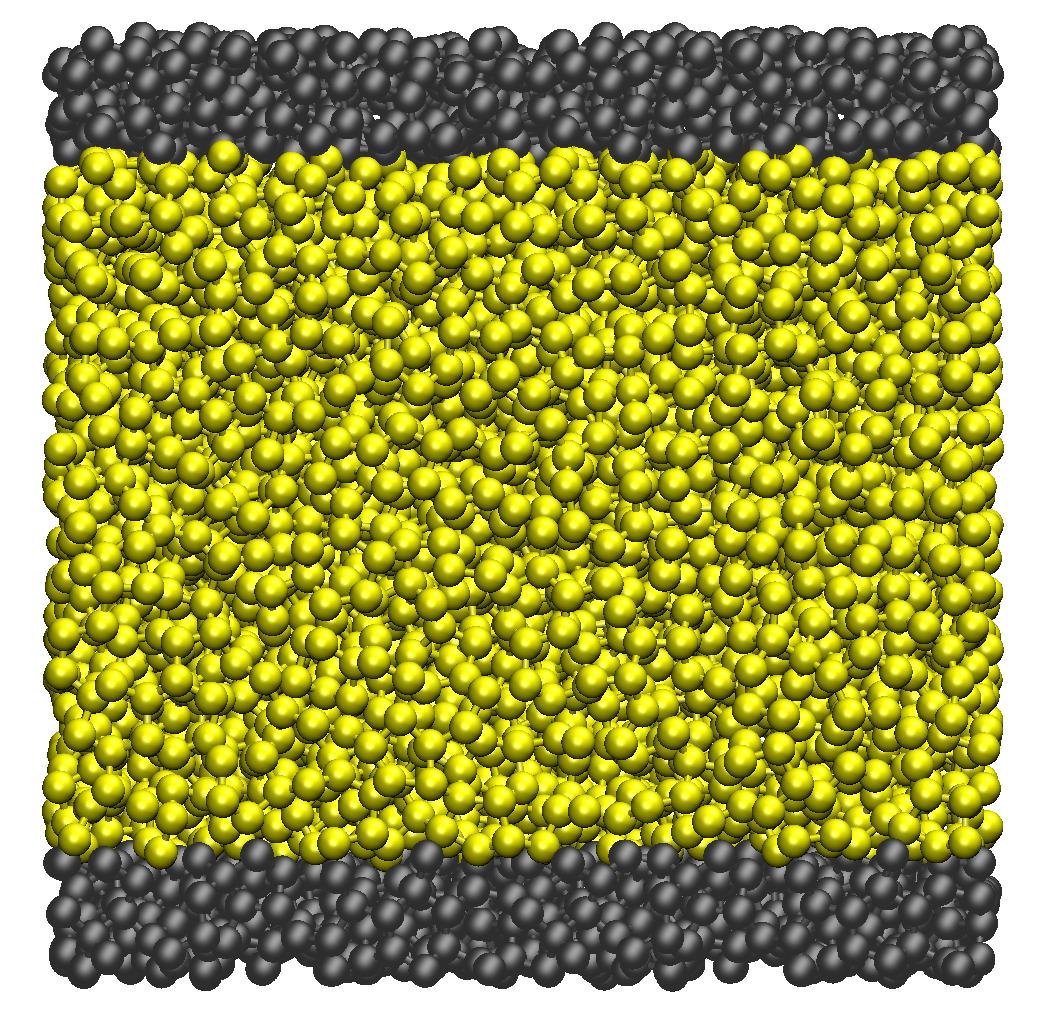}
\caption{(color online) A typical snapshot of the (8000-atom) \aSi~model; the mobile particles are in the center region between two parallel walls (see text for details).}
\label{FigSnapShear}
\end{figure}

The method adopted to apply the mechanical shear deformations has been discussed in detail in Ref.\cite{Kerrache} Three regions are defined, as illustrated in Fig.~\ref{FigSnapShear}: an upper and a lower walls (perpendicular to the $y$-axis) used to apply the mechanical shear deformations, and a central region, with unconstrained mobile particles. The particles in the lower wall are fixed while those in the upper wall move as a whole in the shear direction ($x$) at fixed velocity $v_s$ (shear velocity). The distance between the two walls ($\simeq 42$~\Ang) is kept constant during the simulations. The equations of motion for the mobile particles are integrated using the velocity-Verlet algorithm with a time step of 1~fs. Periodic boundary conditions are imposed in the $x$ and $z$ directions, while they are fixed in the $y$ direction. The thickness of the walls ($\simeq$6~\Ang) is chosen such as they each contain 1000 particles; the mobile region thus contains 6000 particles. The walls are made out of material from the \aSi~matrix, so that their structure is similar to that of the bulk before shear is applied; this choice ensures that no bias towards crystal growth or structural modifications can be induced by the wall layers.

To follow the system's response to shear deformations, we compute the radial distribution function (RDF), the potential energy difference (PED) and the fraction of defects (5-fold coordinated atoms) between the deformed system and the initial non-sheared model.\cite{Kerrache} To follow a possible crystallization process, we employ a topological description based on the ring structure of the two small crystalline building blocks that characterize the diamond and the hexagonal structures.\cite{Beaucage}

\section{\label{res}Shear Deformation Results}

\subsection{\label{crystal}Effects of shearing as a function of temperature}

We have shown, in a previous study,\cite{Kerrache} that high shear rates (relative to the temperature-dependent relaxation rate) induce strong and inhomogeneous local strains in the plastic regime; to allow the system to adapt to the shear deformations, it is therefore necessary to drive it as slowly as possible. On the basis of this work, we select a shear velocity of $10^{-5}$~\Vshear, the slowest rate that allows the system to reach the plastic regime at all temperatures studied in the time scale accessible by MD simulations.

\begin{figure}[ht]
\includegraphics[width=\figurewidth]{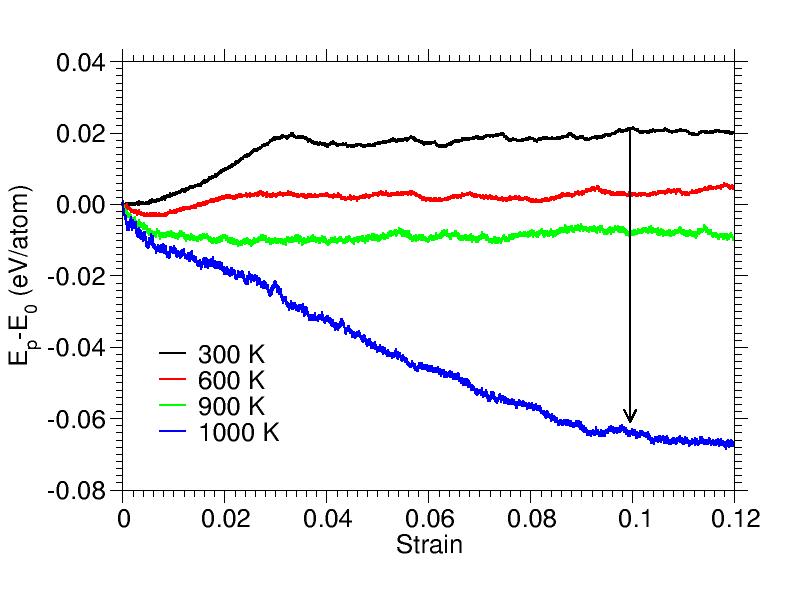}
\caption{(color online) PED between sheared and non-sheared systems at temperatures 300, 600, 900, and 1000~K. The arrow indicates increasing temperature.}
\label{FigEpotTempVs5}
\end{figure}

Simulations were performed at four different temperatures (300, 600, 900 and 1000~K) and were stopped when the total strain reached 12~\%, which is sufficient to induce plastic deformations of amorphous silicon.\cite{Kerrache} The total simulation time of each run is about 500~ns. Figure~\ref{FigEpotTempVs5} shows the evolution of the PED between sheared and non-sheared system as a function of imposed strain for the different temperatures considered. While a systematic increase of the PED was observed at all temperatures when the deformations are driven with high shear velocity,\cite{Kerrache} Fig.~\ref{FigEpotTempVs5} presents a more complex picture with three well-defined regimes: an overall energy increase for the two lowest temperatures, slight relaxation at 900~K, and clear energy decrease at 1000~K.

More precisely, for the two lowest temperatures, 300 and 600 K, the potential energy increases by 0.02 and 0.004~eV/atom, respectively, from the initial equilibrium state to the steady plastic regime. This behavior is characteristic of strain-induced disordering, in agreement with the results discussed by us in Ref.~\onlinecite{Kerrache}. At 900~K, there is no peak associated with the elastic to plastic transition; we observe rather a drop in energy at low strain to a plateau 0.005~eV/atom below the initial value. Shearing thus allows the system to reach a lower-energy state --- still amorphous --- as had been observed, for example, by Isner and Lacks.\cite{Lacks} At 1000~K, finally, the system's response to the deformation is qualitatively different: the energy falls almost linearly up to a strain of about 10~\%, reaching a value about 0.06~eV/atom below the initial state. The energy continues to fall, albeit at a slower rate, for strains up to 12~\%. Clearly, the energy change resulting from the application of shear is the manifestation of strong structural modifications in the network.

\begin{figure}[ht!]
\begin{tabular}{cc}
\centerline{\includegraphics[width=\figurewidth]{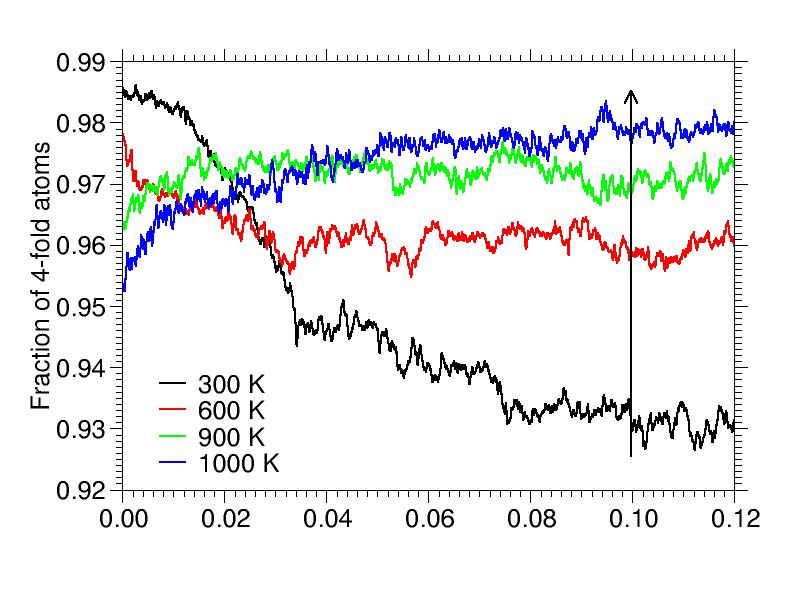}} \\
{~~~~~~~~~~~~~~~~~~~~~~~~~~~~} \\
\centerline{\includegraphics[width=\figurewidth]{figure3a.jpg}}
\end{tabular}
\caption{(color online) 4- (top panel) and 5-fold (bottom panel) coordinated atoms as a function of strain at temperatures 300, 600, 900, and 1000~K. A cutoff of 2.8~\Ang~is used for d
efining which atom are included in the nearest-neighbour count. The arrow indicates increasing temperature.}
\label{FigCordTempVs5}
\end{figure}

To understand the microscopic origin of these changes, it is useful to examine the behavior of some structural parameters. We present, in Fig.~\ref{FigCordTempVs5}, the variation as a function of strain of the proportion of 4- and 5-fold coordinated atoms for the four temperatures considered. We have demonstrated, in a previous article,\cite{Kerrache} that the increase in potential energy of \aSi~resulting from shear is related to the increase in the number of defects, mainly 5-fold coordinated atoms that are considered as ''liquid-like''. As suggested in Refs.~\onlinecite{Argon} and~\onlinecite{Demkowicz}, these defective liquid-like atoms are directly associated with plastic deformations in amorphous silicon.

Fig.~\ref{FigCordTempVs5} demonstrates, here also, a direct correlation between changes in the potential energy and defects created by the shear deformation. Our simulations show that the number of 5-fold atoms (liquid-like] atoms first increases as a function of shear at low temperatures, but decreases at 900~K and 1000~K. At equilibrium (zero strain), the fraction of 5-fold defects increases with temperature, from 1.5~\% at 300 K to 4.9~\% at 1000~K, as can be seen in Fig.~\ref{FigCordTempVs5}. Upon shearing, in contrast, the system crosses over to the opposite situation: the relative number of defects decreases as temperature increases. Thus, from almost 7~\% at 300 K for maximum strain --- the system is unable to relax the defects resulting from the shear ---, it drops to 4~\%, at 600K and to 1.6~\% at 1000~K. As with the PED, the 900~K simulation shows the least impact under shear, with the proportion of defects dropping from 3.5~\% to about about 2.5~\% at a 12~\% strain.

\begin{figure}[ht!]
\includegraphics[width=0.8\figurewidth]{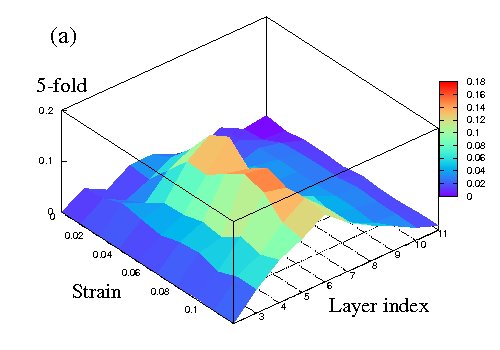}
\includegraphics[width=0.8\figurewidth]{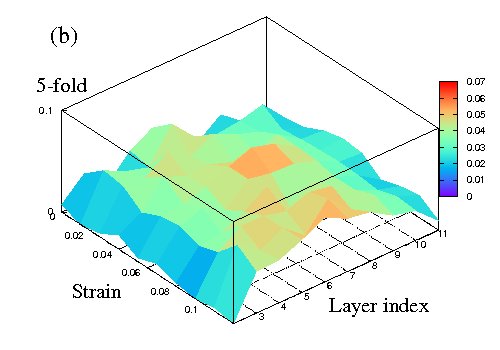}
\includegraphics[width=0.8\figurewidth]{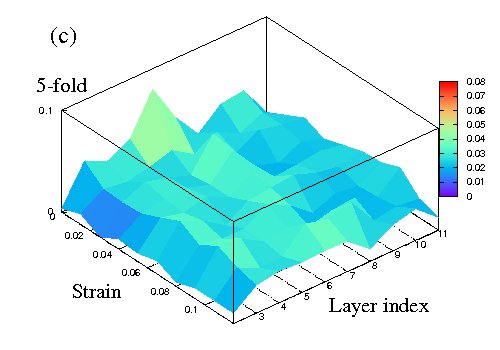}
\includegraphics[width=0.8\figurewidth]{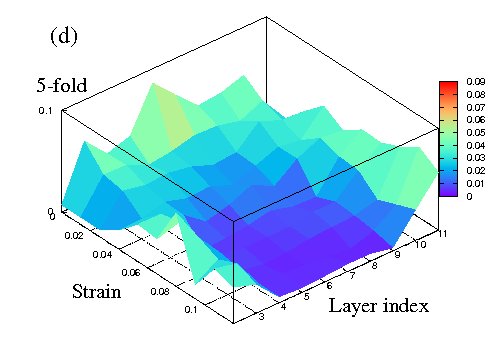}
\caption{(color online)
Fraction of 5-fold coordinated atoms at 300~K (a), 600~K (b), 900~K (c) and 1000~K (d). The configurations are divided into 12 layers along the $y$ direction; layers 1 and 12 are not shown as they correspond to the frozen walls.}
\label{Label3D}
\end{figure}

To characterize the response of amorphous silicon to shear deformations, we show in Fig.~\ref{Label3D} a 3D representation of the distribution of defects as a function of strain and layer index (distance from the edges of the simulation cell --- see caption). This figure reveals that, in the plastic regime, the defects are mostly concentrated in a few layers at 300 K, but that this distribution widens as temperature is raised and becomes uniform at 900~K. While the defect concentration is similar between 900 and 1000~K, we note that the distribution are qualitatively different and defects are concentrated at the interface with the rigid walls for the highest temperature (Fig.~\ref{Label3D} (d)). The decrease of the defect
fraction at 900~K is consistent with the small relaxation observed in the PED. At 1000~K, the small proportion and flat distribution of defects over several layers indicate significant changes in the structure of the sheared material. As we discuss below, this behavior is associated with crystallization spreading over a large region about the center of the system. Since the walls are frozen in the \aSi~structure, the extent of crystallization is limited by the presence of the walls and liquid-like defects are likely to occur at the amorphous-crystal interface.

\begin{figure}[ht]
\includegraphics[width=\figurewidth]{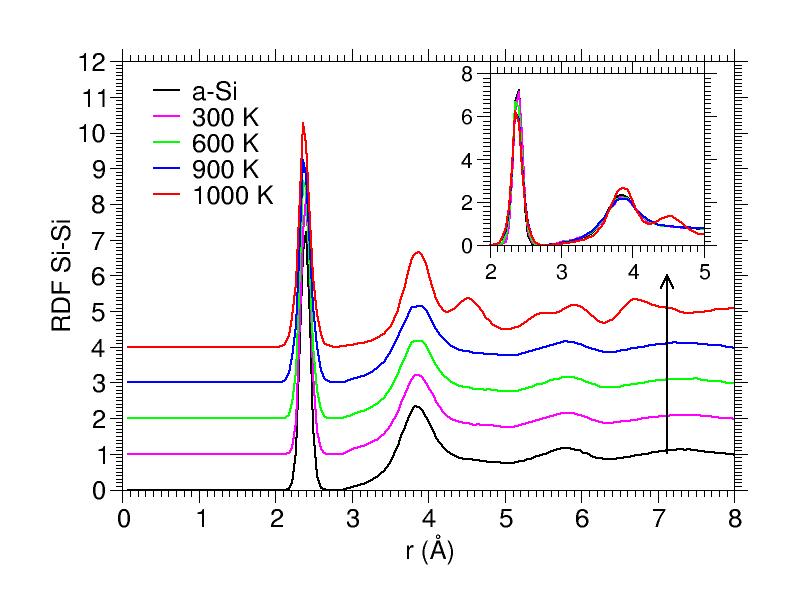}
\caption{(color online) Radial distribution function of the system in the non-sheared state and at 12~\% strain for the four temperatures considered; the curves are shifted for clarity in the main diagram and overlaid in the inset.}
\label{FigRdfCrystTempVs5}
\end{figure}

To obtain a first characterization of the structural significance of these changes, we examine the radial distribution function (RDF) of these four models at maximum strain (12~\%) and compare it with that of unstrained \aSi~at 300~K (Fig.~\ref{FigRdfCrystTempVs5}). At 12~\% strain, the energy and the defect density are saturated in all cases and are thus representative of the steady state plastic regime.

For temperatures 300, 600, and 900 K, the RDFs are very similar and resemble closely that of relaxed \aSi. The positions of the first- and second-neighbor peaks coincide, as evident in the inset. The main manifestation of shearing is the formation of a new and small structure between the first and second peaks in the RDF. This is associated with 5-fold coordinated atoms and was discussed in details in Ref.~\onlinecite{Kerrache}; it is strongest at the lowest temperatures and essentially disappears at 900~K, consistent with the results of Fig.~\ref{FigCordTempVs5}. Interestingly, at this temperature, the shearing counterbalances the effects of thermal disorder, yielding an amorphous state with fewer coordination defects (lower energy) than the original model. The amorphous model is thermodynamically meta-stable and the combination of the shearing and thermal effects brings the system to a more relaxed structure with fewer defects and low energy.

At 1000~K, now, a completely different behavior is observed: the RDF is qualitatively different from the low-temperature ones, exhibiting much better-defined peaks, corresponding in fact to the crystalline state: closer examination of the system reveals that it is actually crystallizing under the yoke of shear. However a complete crystallization of the system cannot be reached since it is limited by the presence of the walls which remain amorphous during the whole shearing process.

\subsection{\label{extent}Extent of crystallization}

\begin{figure}[ht!]
\includegraphics[width=\figurewidth]{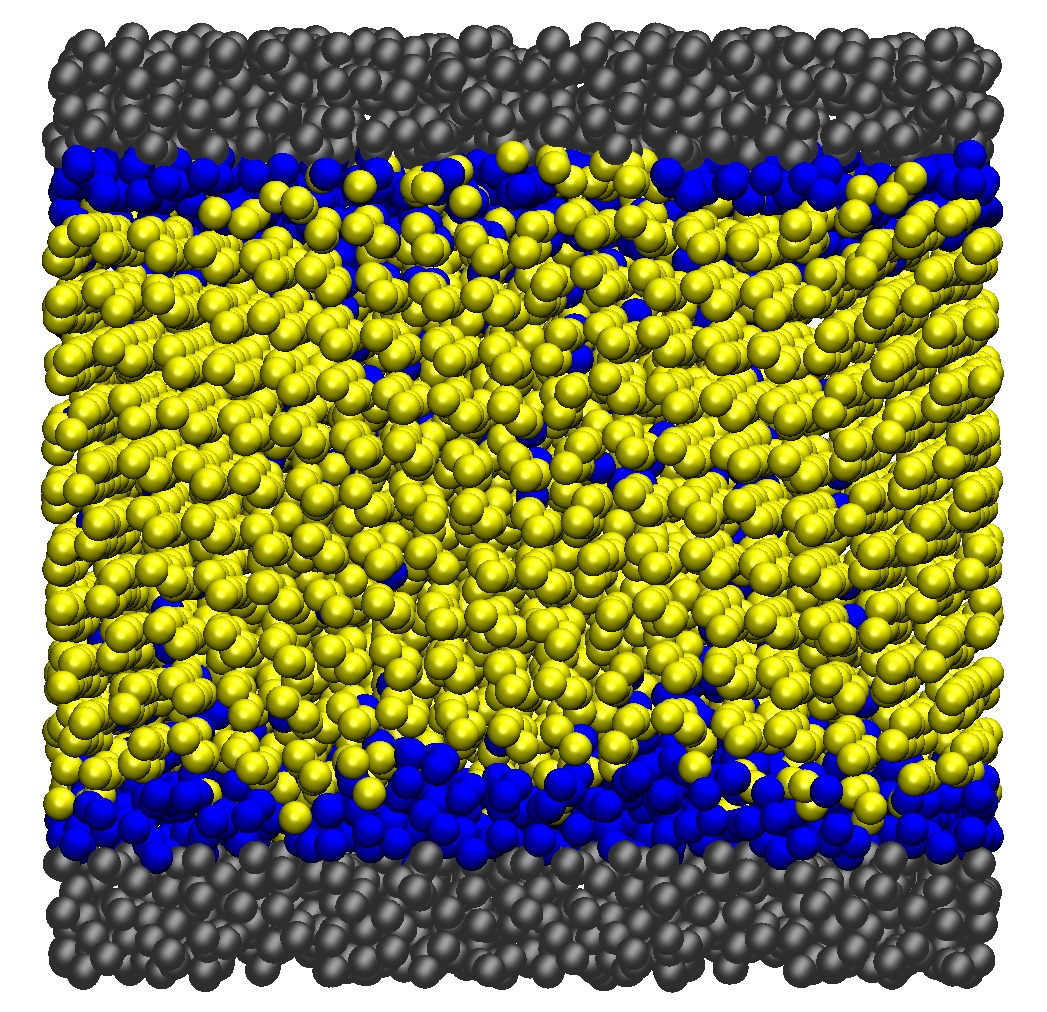}
\includegraphics[width=\figurewidth]{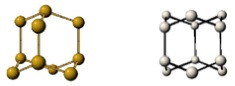}
\caption{(color online) Snapshot of the system showing the distribution of crystalline particles, in light/yellow, at 1000~K; amorphous and liquid-like particles are in dark/blue, and the rigid walls are in black. The bottom panel shows the blocks used to characterize the local crystalline order: \emph{Wurtzite} (hexagonal) structure (right) and diamond structure (left); see text for details.}
\label{FigCrystTempVs5}
\end{figure}

The extent to which crystallization takes place is best understood, and quantified, in terms of an order parameter. For this purpose, we use a topological characterization based on two-ring structures (''blocks'') associated with local crystalline order;\cite{Nakhmanson, Beaucage} they are displayed in Fig.\ref{FigCrystTempVs5} (bottom panel). These blocks represent the smallest 3D rigid structures that can be extracted from the tetrahedral lattice. The first one derives from the \emph{Wurtzite} (hexagonal) structure and contains 12 atoms; the second is extracted from the diamond structure and is made up of 10 atoms assembled as four 6-fold rings back to back. A similar approach was used, for example, in dense-packed systems by Mokshin and Barrat.\cite{Mokshin-08, Mokshin-09}

We present in Fig.~\ref{FigCrystTempVs5} a snapshot of the system undergoing crystallization; this corresponds to a configuration obtained at the end of the simulation at 1000~K and a total strain of 12~\%. Clearly, the amorphous particles (blue) are concentrated near the walls (dark) while atoms within a crystalline environment occupy the bulk region (yelow). The high concentration of the amorphous particles at the interfaces between the walls and the bulk region is due to the presence of the walls. Let us mention that the structure of the walls remain the same during shear deformations. Therefore the crystallization of amorphous silicon is limited by the amorphous nature of the walls. This situation is similar to that found in solid-liquid or amorphous-crystal interface studies. The interface is extended over several layers, as for example see the Ref.~\onlinecite{Kerrache08}. 

\begin{figure}[ht]
\includegraphics[width=\figurewidth]{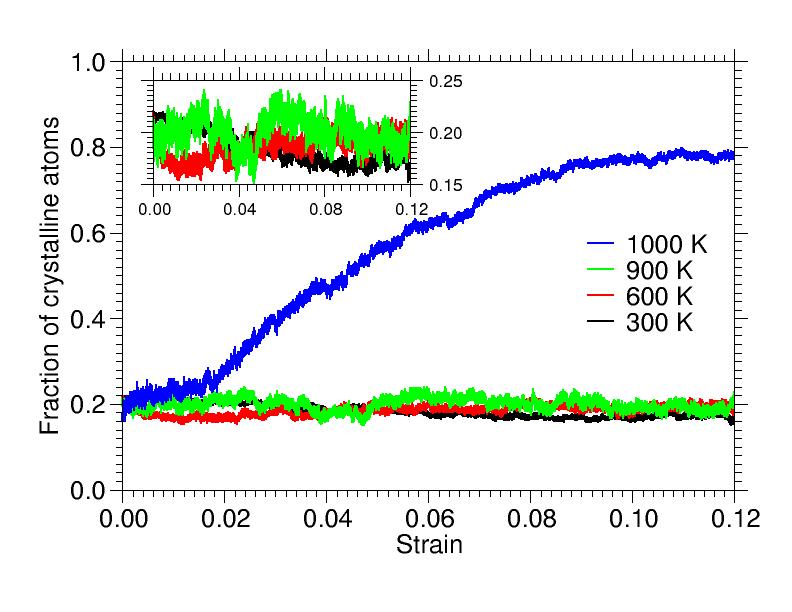}
\caption{(color online) Fraction of crystalline particles versus strain for the four temperatures considered.}
\label{FigVelCrystTempVs5}
\end{figure}

We present in Fig.~\ref{FigVelCrystTempVs5} the fraction of atoms which can be tagged as crystalline within the central region of the system as a function of strain for the four temperatures considered. At zero strain, about 20~\% of all atoms can be associated with a wurtzite or diamond structure at all temperatures. Since the crystal-like particles are distributed uniformly across the simulation box, these are associated with local topological fluctuations and do not represent a crystalline phase; as discussed in Ref.~\onlinecite{Beaucage}, the specific fraction of these crystal-like atoms in a well-relaxed amorphous network actually depends on the details of the interatomic potential.

Consistent with the results presented in the previous section, the proportion of crystalline atoms remains approximately constant at the three lowest temperatures considered (300, 600, 900 K), while fluctuations get larger as temperature increases (see inset). These fluctuations indicate that, by increasing temperature, a large fraction of atoms see their local topological environment change between amorphous and crystalline state. The wider distribution of the slip planes; slippage is concentrated in a few layers at low temperature, as was shown in Fig.~\ref{Label3D}, and involves a larger fraction of the box as the temperature rises. 

At 1000~K, however, we observe a qualitative change. The fraction of crystalline particles increases steadily --- slowly for strains $\leq 2$~\%, then more rapidly up to $\simeq$10~\%, where it reaches a plateau close to 80~\%. Such a three-stage curve is similar to that observed in the crystallization of silicon from the liquid phase (see for example Fig.~4 of Ref.~\onlinecite{Beaucage}), suggesting a strain-induced nucleation process leading to crystallization, limited only by the presence of amorphous walls and grain boundaries. Complete crystallization however cannot be expected, especially close to the interfaces with the walls.

\section{\label{discuss}Discussion}

While several experiments have been conducted to understand the micro-mechanisms of plasticity in metallic glasses, precise understanding of the atomic processes involved in shear propagation is far from being complete. In particular, several experiments suggest the possibility of deformation-induced crystallization, as well as nano-void formation.\cite{Lee, Kim, Li, Jiang, Bhowmick, Lee06} Likewise, recent work on single-component Lennard-Jones systems has shown that shearing can lead to crystallization from the disordered state.\cite{Mokshin-08, Mokshin-09} However, this has not yet been confirmed for more complex, and more stable, disordered systems such as amorphous silicon. Indeed, this material is known to be very stable at low temperature; it can nevertheless crystallize at temperatures well below the melting point and numerous studies have focused on crystallization directly from the liquid phase\cite{Beaucage} or at the liquid-crystal or amorphous-crystal interfaces.\cite{Buta, Krzeminski, Bernstein98, Mattoni, Lewis96} Here, we have shown that, when external strain is imposed by a very low shear velocity, three different regimes are observed: (i) disordered, (ii) annealed and (iii) crystallized. The first regime is systematically observed at low temperatures or at high shear velocity (see Ref.~\onlinecite{Kerrache} for more details). The second and third regimes are clearly observed at 900 and 1000~K, respectively.

As we have discussed previously,\cite{Kerrache} when the shear rate is faster than thermal relaxation, the steady-state plastic regime involves the creation of additional 5-fold, liquid-like coordinated atoms that facilitate the constant reorganization of the network; the system is thus ''more disordered'' compared to the initial model. This behavior has also been characterized for various other silicon classical potentials by Talati \emph{et al.}\cite{Talati}

The second regime prevails when thermal effects are sufficiently large to counterbalance the shear deformations. In this case, the applied strain allows the system to sample more efficiently the energy landscape, finding new basins with low-energy states that can be reached due to the available thermal energy.\cite{Lacks} This is what we observe at 900~K for all quantities studied: energy per atom (Fig.~\ref{FigEpotTempVs5}), defect density (Fig.~\ref{FigCordTempVs5}), radial distribution function (Fig.~\ref{FigRdfCrystTempVs5}) and proportion of crystalline particles (Fig.~\ref{FigVelCrystTempVs5}). From local to global properties, shearing at this high temperature leads to deeper annealing. Interestingly, even though considerable rearrangements take place, the available thermal energy is not sufficient to allow nucleation of the crystal and the phenomenon is more akin to standard thermal annealing used for producing ''good quality'' glasses and amorphous materials.

At 1000~K, however, the combination of temperature and shearing leads to crystallization. The evolution of the order parameter shown in Fig.~\ref{FigVelCrystTempVs5} follows closely that observed for the nucleation from the liquid phase\cite{Beaucage}: first, small nuclei appear with a slow increase of the order parameter; this is followed by the rapid growth of the largest nuclei until the grains touch and the system becomes poly-crystalline. Remarkably, the time scale over which crystallization takes places from the liquid phase is more than a 100 times slower than for shearing; this is likely due to the fact that diffusion is much slower in the amorphous than in the liquid state, and so crystals cannot assemble as quickly.

The validity of our results can be established to some degree by noting that they are consistent with the corresponding situation in such systems as amorphous alloys based on iron and nickel.\cite{Ratushnyak} Some experiments have shown, for example, that an external driving force can induce crystallization in jammed systems that can be considered as models for disordered materials.\cite{Mokshin-08, Mokshin-09} Vibrations,\cite{Vanel, Daniels} shear oscillations,\cite{Nicolas, Mueggenburg} and steady shear\cite{Tsai} have been shown to induce crystallization in granular systems of spherical particles, and shear oscillations induce crystallization in colloidal glasses.\cite{Haw}

We note also that crystallization was not observed by Talati \emph{et al.}~\cite{Talati} in their MD simulations using both the SW and the Tersoff potential; this is likely related to the fact that they have considered relatively low temperatures. We speculate that the details of the crystallization process --- temperature range, shear rate, etc. --- are potential dependent, but the physics is not.

In contrast to single-component Lennard-Jones, amorphous silicon is a metastable phase that can be annealed experimentally without crystallization; this allows us to observe a much slower crystallization process,  emphasizing the similarities with classical nucleation. This suggests that this phenomenon could be more much generic that initially thought and adds weight to previous observation relating shear and temperature.\cite{Chen, Rottler, Tarumi, Robertson, Mattoni, Kim, Jiang, Beaucage, Ratushnyak} This relation is not perfect, clearly, and further studies are necessary to fully understand how one can integrate external deformations into standard nucleation theory.

\section{\label{conclude}Conclusion}

In summary, we have investigated the behavior of amorphous silicon subject to slow shear deformations using classical MD simulations with the EDIP potential. While we found a systematic increase of the PED and the defects fraction at low temperatures or at high shear velocities, the shear deformations, when induced by low shear velocities, lead to three different regimes that emerge as a function of temperature: increased disorder for temperatures below $\simeq 900$~K, enhanced relaxation around 900~K, and crystallization at $\simeq 1000$~K. The crystallization of amorphous silicon appears to follow classical nucleation. While the exact parameter space for these three regimes to develop certainly depends on the choice of interatomic potential, our results, which are consistent with a number of other observations, appear to be robust and open a new approach for modifying and manipulating glasses, and even controlling the formation of \emph{crystalline} nano-structures
inside disordered matrices.

\begin{acknowledgments}

This work has been supported by grants from the \textit{Fonds Qu\'{e}b\'{e}cois de la Recherche sur la Nature et les Technologies} (FQRNT) and the Natural Sciences and Engineering Research Council of Canada (NSERC). The simulations were done on the computers of the \textit{R\'eseau Qu\'eb\'ecois de Calcul de Haute Performance} (RQCHP), whose support is gratefully acknowledged.

\end{acknowledgments}

\end{document}